\documentclass[twocolumn,showpacs,prl,aps,superscriptaddress]{revtex4-1}
\usepackage{graphicx}
\usepackage{epstopdf}
\usepackage{color}
\usepackage{amsmath}
\usepackage{subfigure}
\usepackage{float}

\mathchardef\mhyphen="2D
\newcommand{\TMB}{Ti$_{4}$MnBi$_{2}$}

\begin{document}
\title{Correlations and incipient antiferromagnetic order within the linear Mn chains of metallic Ti$_4$MnBi$_2$}

\author{Abhishek Pandey}
\altaffiliation{abhishek.pandey@wits.ac.za}
\affiliation{School of Physics, University of the Witwatersrand, Johannesburg, Gaugent 2050, South Africa}
\affiliation{Department of Physics and Astronomy, Texas A$\&$M University, College Station, Texas 77843, USA}
\author{Ping Miao}
\author{M. Klemm}
\author{H. He}
\affiliation{Department of Physics and Astronomy, Texas A$\&$M University, College Station, Texas 77843, USA}
\author{H. Wang}
\author{X. Qian}
\affiliation {Department of Materials Science and Engineering, Texas A$\&$M University, College Station, Texas 77843, USA}
\author{J. W. Lynn}
\affiliation{NIST Center for Neutron Research, National Institute of Standards and Technology, Gaithersburg, Maryland 20899, USA}
\author{M. C. Aronson}
\altaffiliation{meigan.aronson@ubc.ca}
\affiliation{Department of Physics and Astronomy, Texas A$\&$M University, College Station, Texas 77843, USA}
\affiliation {Department of Materials Science and Engineering, Texas A$\&$M University, College Station, Texas 77843, USA}
\affiliation {Department of Physics and Astronomy, University of British Columbia, Vancouver V6T 1Z4, British Columbia, Canada}
\affiliation {Stewart Blusson Quantum Matter Institute, University of British Columbia, Vancouver V6T 1Z4, British Columbia,
Canada}
\date{\today}

\begin{abstract}
We report measurements on \TMB, where a crystal structure involving linear chains of Mn ions suggests one-dimensional magnetic character. The electrical resistivity is metallic, consistent with the results of electronic structure calculations that find a robust Fermi surface albeit with moderate electronic correlations. Curie-Weiss fit to the magnetic susceptibility finds that the Mn moments are in the low-spin $S = 1/2$ configuration. Neutron diffraction measurements detect weak antiferromagnetic order within the Mn chains, with further evidence for the small staggered moment coming from the entropy associated with the ordering peak in the specific heat as well as from the results of spin-polarized electronic structure calculations. The antiferromagnetic moments are apparently associated with the $d_{x^{2}-y^{2}}$ and $d_{xy}$ orbitals of Mn while the remaining Mn orbitals are delocalized. Strong quantum fluctuations, possibly related to an electronic instability that forms the Mn moment or to the one-dimensional character of \TMB, nearly overcome magnetic order.
\end{abstract}

\pacs{71.27.+a, 75.10.Pq, 75.25.-j}
\maketitle

\section{Introduction}
Low dimensional magnetic systems continue to excite since strong quantum fluctuations suppress magnetic order, showcasing novel quantum effects at the lowest temperatures~\cite{Giamarchi-2004}. Of particular importance is the extensive body of theoretical results on one-dimensional (1D) systems that can be directly tested by  experiments like inelastic neutron scattering~\cite{lake2009,mourigal2013} and resonant inelastic x-ray scattering~\cite{schlappa2018}. To date, most experiments have been carried out on spin $S = 1/2$ systems like KCuF$_3$ \cite{Hirakawa-1967, Lake-2005, Lee-2012, Pavarini-2008}, SrCuO$_2$ \cite{Kim-1996}, Sr$_2$CuO$_3$ \cite{Schlappa-2012} and CuSO$_{4}$.5D$_2$O~\cite{mourigal2013}, which are strongly correlated insulators. Inelastic neutron scattering experiments demonstrate that the fundamental excitations are not the spin waves expected in three dimensional (3D) systems, but rather spinons, demonstrating that the electron spin and charge have become separated~\cite{Wu-2016,Jompol-2009,Lake-2005}.

It is less clear what happens to this scenario when electronic correlations are not strong enough to sustain an insulating gap. The first examples of metallic spin chains were realized in the organic conductors~\cite{kanoda2011,jerome2012,ardavan2012}, where the ratio of the Coulomb interaction $U$ and the electron bandwidth 4$t$ are approximately equal, $\it{i.e.}$ $U/4t \simeq$ 1. Angle-Resolved Photoemission Spectroscopic (ARPES) measurements found that the fundamental exitations are gapless spinons and holons in the organic conductor tetrathiafulvalene tetracyanoquinodimethan (TTF-TCNQ) \cite{claessen2002}, confirming theoretical expectations that 1D metals are Tomonaga-Luttinger liquids~\cite{Giamarchi-2004}. Although electronic interactions and interchain coupling can be sufficiently strong in organic conductors~\cite{kanoda2011,jerome2012,ardavan2012} to drive many-body ground states like superconductivity, magnetic order or even metal-insulator transitions, these materials are mechanically fragile and are only available in small quantities. As a result, it is problematic to utilize the most powerful spectroscopic tools, in particular inelastic neutron scattering measurements, on these materials. There remains a pressing need to identify new families of compounds with metallic spin chains where the impact of intermediate electronic interactions may be assessed. In particular, the near absence of suitable 1D materials has meant that the interplay of magnetic correlations  and Kondo compensation of the underlying moments, central to the rich physics of the 3D heavy fermions, remains virtually unexplored by experiments in  their 1D counterparts~\cite{sikkema1997,tsunetsugu1997, tsvelik2015,schimmel2016,khait2018}. We report here measurements on an unexplored metallic system \TMB\ that suggest the 1D character implied by its unique crystal structure may control its fundamental properties.

\TMB\ has a remarkable crystal structure~\cite{Richter-1997, Rytz-1999} with well separated perfectly linear chains of Mn ions. Electrical resistivity measurements demonstrate that \TMB\ is a metal and electronic structure calculations confirm that there is a robust Fermi surface with moderate mass enhancement. The Mn moments are reasonably localized and are in the low-spin $S = 1/2$ configuration. Antiferromagnetic (AFM) order is observed near 2~K due to weak interchain coupling. However, the ordered moment is much smaller than it would be expected in an insulator, indicating that quantum fluctuations are likely quite strong in \TMB, possibly due to its 1D character or proximity to an electronic instability that produces a magnetic moment. Our measurements suggest that $d$-electron based intermetallic compounds like \TMB\ are likely a new productive direction to explore, in which robust magnetic chains are embedded in metallic hosts.

\begin{figure*}
\includegraphics[width=6in]{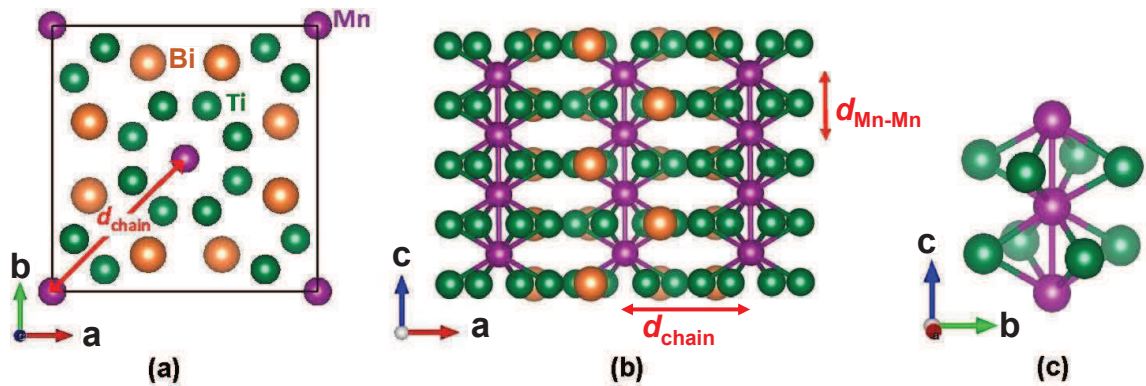}
\caption{Crystal structure of tetragonal \TMB. (a) A view of the crystal structure in the basal plane, depicting the
central Mn ion surrounded by an inner ring of eight Ti ions and an outer ring of eight Bi ions. The respective rings of Ti and Bi ions are formed by two sets of four coplanar ions occupying the corners of square-planer lattices that are separated by $d=c/2$ along the $c$-axis and are at an angle of 45$^{\circ}$ with respect to each other. Extending along the $c$-axis,  these rings constitute tube-like structures of Ti and Bi that surround the linear chains of Mn-ions. (b) A view of the two unit cells of \TMB\ stacked on the top of each other along the $c$-axis. It is evident from (a) and (b) that the Mn-ions form linear chains running along the $c$-axis of the tetragonal unitcell. The interchain distance between the two nearest Mn ions within the $ab$-plane is $d_{\rm chain} = a/\sqrt{2} = 7.4208(3)$~\AA\ and the intrachain distance between the two neighboring Mn-ions is $d_{\rm Mn-Mn} = c/2 = 2.4930(1)$~\AA. (c) The Mn-ions centered in a square-antiprism coordination polyhedra of Ti-ions. Two nearest neighbors Mn-ions capping the respective square faces are also shown.}
\label{fig:Structure}
\end{figure*}

\section{Experimental and Computational Details}

Single crystals of Ti$_4$MnBi$_2$ were grown using the solution growth technique. High purity Ti (99.99\%), Mn (99.98\%) and Bi (99.999\%) in the molar ratio of 4:2:10 were placed in an alumina crucible and sealed inside an evacuated quartz tube. The assembly was heated to 1150 $^{\circ}$C in 10~h, kept there for 30~h, and then cooled down to 800 $^{\circ}$C in 100~h. At this temperature the excess Bi-flux was decanted and needle like crystals of the typical size $0.5\times0.5\times3$~mm$^3$ were obtained.

Structural characterization of the compound was performed using single crystal and powder x-ray diffraction (XRD). The single crystal diffraction intensity data were collected at room temperature using a Bruker D8 Quest diffractometer with Mo-$K_{\alpha}$ radiation ($\lambda = 0.71073$~\AA). Data integration/reduction and an absorption correction using a multi-scan method were performed with the Bruker APEX2 software suite. The structure was solved by the direct method and refined by a full matrix least-squares method against F2 using the SHELX package. Room temperature powder XRD data were obtained on a few crushed crystals using a Bruker D8 Advance Powder Diffractometer employing Cu-$K_{\alpha}$ radiation. Rietveld refinement was carried out using the FullPROF \cite{Carvajal-1993} package, keeping the occupancies at each lattice site fixed at $100\%$. A satisfactory refinement of the room temperature powder XRD data confirms the single phase nature of the material (Fig.~S1, Supplementary Materials) as well as the integral site occupancies and the absence of any measurable site disorder. A similar conclusion was reached by refinements of the single crystal
XRD data.

Magnetic measurements at different temperatures $0.45~K \leq T \leq 300~K$ and magnetic fields $H\leq 70$~kOe were carried out using a Quantum Design Magnetic Properties Measurement System fitted with a $^{3}$He option as well as a Vibrating Sample Magnetometer installed in a Quantum Design Physical Properties Measurement System (PPMS). Measurements of the heat capacity $C_{\rm p}(T)$ and $c$-axis electrical resistivity $\rho_{\rm c}(T)$ were performed using the $^{3}$He option of the PPMS. In the latter experiment, electrical contacts were made using silver-filled epoxy and the measuring current was 500 $\mu$A\@.

Neutron powder diffraction data were collected on a 4~g powder sample prepared by crushing the single crystals of \TMB. The data were obtained with a neutron wavelength of 2.359~\AA\ using the double focusing triple-axis spectrometer BT-7 at NIST Center for Neutron Research \cite{lynn2012}, which is equipped with position sensitive detectors and a dilution refrigerator with a base temperature of 0.05 K. Rietveld refinement of these data was performed using the software Z-Rietveld \cite{Oishi-2009, Oishi-2012}. The residual value of $R_{\rm wp}$ was below 10\%. In addition, single crystal neutron diffraction measurements were performed on BT-7 using ~2~g aligned single crystals of \TMB\ mounted in a close-cycle refrigeration unit that obtains temperatures below 2.5~K. Uncertainties, where indicated, represent one standard deviation.

The electronic structure of \TMB\ was calculated using first-principles density-functional theory~\cite{hohenberg1964,kohn1965} as implemented in the Vienna Ab initio Simulation Package (VASP)~\cite{kresse1996} with the projector-augmented wave method~\cite{blochl1994}. We employed the Perdew-Burke-Ernzerhof exchange-correlation functional~\cite{perdew1996}, a plane-wave basis with an energy cutoff of 200 eV, and a Monkhorst-Pack k-point sampling of 4$\times$4$\times$4 for Brillouin zone integration~\cite{klimes2010} as well as spin-orbit coupling. A Hubbard $U$ correction~\cite{dudarev1998} of $U_{\rm eff} = 3.9$~eV was applied to treat the strong Coulomb interactions of localized $d$ electrons on the Mn atoms.

\section{Results}
\subsection{Structural Properties}

\begin{table}
\caption{Room temperature crystallographic parameters estimated from x-ray powder diffraction data of tetragonal \TMB, which crystallizes in a V$_4$SiSb$_2$-type structure with $I4/mcm$ (No. 140) space group symmetry and $Z = 4$ formula units/cell.}

\label{Table:StructureData}
 \begin{ruledtabular}
		\begin{tabular}{l l c c c}
		 Atom & Wyckoff  & $x$ & $y$ & $z$ \\
		      & Position &     &     &     \\
	 \hline
			Ti & $16k$ & 0.0804(3) & 0.1977(4) & 0 \\
			Mn & $4a$ & 0 & 0 & $\frac{1}{4}$ \\
			Bi & $8h$ & 0.1390(1) & 0.6390(1) & 0 \\
	 \hline
	  {\bf Lattice Parameters} & & & &\\
		$a~({\rm \AA})$ & 10.4946(4) & & & \\
		$c~({\rm \AA})$ & 4.9860(2) & & & \\
		$c/a$ & 0.47510(4) & & & \\
		$V_{\rm cell}~({\rm \AA}^3)$  & 549.14(4) & & & \\
		$d_{\rm Mn\mhyphen Mn}~({\rm \AA})$  & 2.4930(1) & & & \\
		$d_{\rm chain}~({\rm \AA})$  & 7.4208(3) & & & \\
	\end{tabular}
	\end{ruledtabular}
\end{table}

\TMB\ has been previously reported to form in the V$_4$SiSb$_2$-type tetragonal structure (space group: $I4/mcm$, No. 140),
\cite{Richter-1997, Rytz-1999} and the structural parameters obtained from our XRD refinements (Table~\ref{Table:StructureData}) are in good agreement with the published results. The crystal structure of \TMB\ depicted in Fig.~\ref{fig:Structure} emphasizes its 1D character. Linear chains of Mn atoms with a Mn-Mn spacing of $d_{\rm Mn\mhyphen Mn} = 2.4930(1)~{\rm \AA}$ extend along the tetragonal $c$-axis. The Mn chains are encased along the $c$-direction in concentric tubes with an inner tube of Ti atoms and a larger diameter tube of Bi atoms [Fig.~\ref{fig:Structure}(a)]. The distance between Mn chains $d_{\rm chain} = 7.4208(3)$~\AA\ is substantially larger than $d_{\rm Mn\mhyphen Mn}$ [Fig.~\ref{fig:Structure}(b)], suggesting a pronounced one-dimensionality of the Mn subsystem. The Mn-ions are centered in a square-antiprism coordination polyhedra of Ti-ions [Fig.~\ref{fig:Structure}(c)].

\subsection{Electrical Resistivity}
\begin{figure}
\includegraphics[width=3.3in]{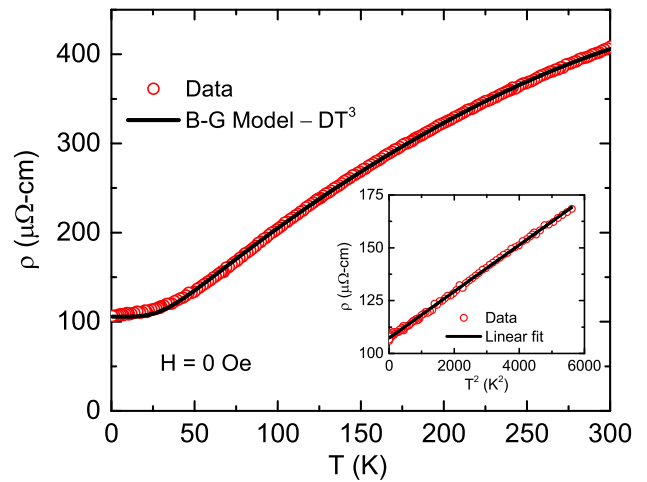}
\caption{Temperature dependence of the electrical resistivity $\rho_{\rm c}(T)$ of \TMB\ measured along the crystallographic $c$-axis. The solid black curve is a fit by the Bloch-Gr\"uneisen model with an additional term $-DT^3$ ascribed to Mott $sd$ scattering (see text). Inset: $\rho_{\rm c}$ plotted as a function of temperature squared $T^2$ for $T \lesssim 75$~K\@. Solid black line is a linear fit to the data.}
\label{fig:Res-1}
\end{figure}

We begin by establishing that \TMB\ is intrinsically metallic. The temperature dependence of the $c$-axis electrical resistivity $\rho_{\rm c}(T)$ measured for temperatures between $0.5-300$~K is presented in Fig.~\ref{fig:Res-1}. Similar results were obtained on multiple crystals, with little variation in the residual resistivity $\rho_{0}$, which was found to vary between $100-120$ $\mu\Omega$ cm. The overall good quality of our single crystals and the lack of significant sample dependence suggest that disorder is unlikely to be solely responsible for the large value of $\rho_0$, implying that other sorts of scattering processes including quantum fluctuations, may be present at $T = 0$. A metallic temperature dependence was found at all temperatures $0.5~K \leq T \leq 300~K$, with $\rho_{\rm c}(T)$ increasing monotonically with increasing temperature. Fermi liquid behavior was found for $T\lesssim~75~K$ [inset, Fig.~\ref{fig:Res-1}], where the resistivity data were fitted by $\rho_{\rm c} = \rho_{0} + AT^2$, with $A = 1.103(3)\times10^{-3}~\mu\Omega$~cm/K$^2$ (Table~\ref{Table:PhysicalProperties}). This value of $A$ is enhanced relative to values found in simple metals and is similar to those found in correlated $d$-electron compounds like La$_{1.7}$Sr$_{0.3}$CuO$_{4}$ and Sr$_{2}$RuO$_{4}$~\cite{jacko2009}. This is our first indication that the conduction electrons in \TMB\ are significantly correlated.

At higher temperatures, $\rho_{\rm c}(T)$ does not approach the linear temperature dependence that is expected in the Bloch-Gr\"{u}neisen model of conduction carrier scattering from acoustic lattice vibrations. A pronounced downward curvature is observed and a satisfactory fitting of the data is obtained by adding a Mott-$sd$ scattering term $-DT^3$ with $D = 2.43(3) \times 10^{-8}~\mu\Omega\,{\rm cm}/{\rm K}^3$ (Fig. 2). This is another piece of evidence for the presence of electronic correlations that originate with $d$-electrons.

\begin{table}
\caption{Parameters deduced from magnetic, thermal and electrical transport measurements as well as from the electronic structure calculations performed on Ti$_4$MnBi$_2$ single crystals; where $C$ is the Curie constant, $\mu_{\rm eff}$ is the effective paramagnetic moment, $\theta_{\rm p}$ is the paramagnetic wise temperature, $J_{\rm c}$ is the nearest neighbor exchange constant along the $c$-axis, $\chi_{0}$ is the temperature independent susceptibility estimated from the Curie-Weiss fit, $\chi_{\rm Pauli}$ is the Pauli susceptibility estimated from the experimental value of the Sommerfeld coefficient $\gamma$, $\beta$ is the coefficient of lattice heat capacity at low temperatures, ${\cal D}_{\gamma}$ (states/eV~f.u.) for both spin directions is the density of states (DOS) at the Fermi level $E_{\rm F}$ deduced from the $\gamma$ value, ${\cal D}_{\rm band}$ (states/eV~f.u.) for both spin directions is the DOS at $E_{\rm F}$ calculated by electronic structure calculations, $\Theta_{\rm D}$ and $\Theta_{\rm E}$ are Debye and Einstein temperatures estimated from heat capacity data, respectively, $u$ is the fraction of Debye contribution to the heat capacity, $A$ is the coefficient of $T^2$ term in of the resistivity data at low temperatures, $D$ is the coefficient of Mott $sd$-scattering term in the resistivity data, $R_{\rm KW}$ (m$\Omega$~cm~mol$^2$~J$^{-2}$K$^2$) is the Kadowaki-Woods ratio and $R_{\rm W}$ is the Wilson ratio.}
\label{Table:PhysicalProperties}
 \begin{ruledtabular}
		\begin{tabular}{l l  }
		 Parameter & Estimated value  \\
	 \hline
			$C~({\rm cm^3~K/mol})$ & 0.41(2) \\
			$\mu_{\rm eff}~(\mu_{\rm B})$ & 1.79(4) \\
			$\theta_{\rm p}~(\rm K)$ & $-9$(3) \\
			$J_{\rm c}$(K) & 18(6)\\
			$\chi_0~(10^{-4}~{\rm cm^3/mol})$ & 4.8(5) \\
			$\chi_{\rm Pauli}~(10^{-4}~{\rm cm^3/mol})$ & 7.8(9) \\
			$\gamma~({\rm mJ/mol~K^2})$ & 57(5) \\
		  $\beta~({\rm mJ/mol~K^4})$ & 0.94(6) \\
			${\cal D}_{\gamma}(E_{\rm F})$~(unit in the caption) & 24(3) \\
			${\cal D}_{\rm band}(E_{\rm F})$~(unit in the caption) & 7.4(1) \\
			$\Theta_{\rm D,~HT}~(\rm K)$ & 381(15) \\
			$\Theta_{\rm D,~LT}~(\rm K)$ & 355(2) \\
			$\Theta_{\rm E}~(\rm K)$ & 110(22) \\
			$u$~(unitless) & 0.73(7) \\
			$A~(10^{-3}~\mu\Omega~{\rm cm}/{\rm K}^2)$ & 1.103(3) \\
			$D~(10^{-8}~\mu\Omega~{\rm cm}/{\rm K}^3)$ & 2.43(3) \\
			$R_{\rm KW}$~(unit in the caption) & $3.4\times10^{-4}$\\
			$R_{\rm W}$~(unitless) & 0.6(2)
	\end{tabular}
	\end{ruledtabular}
	\end{table}

\subsection{Heat Capacity and Magnetic Entropy}
\label{sec:HC}

\begin{figure}
\includegraphics[width=3.0in]{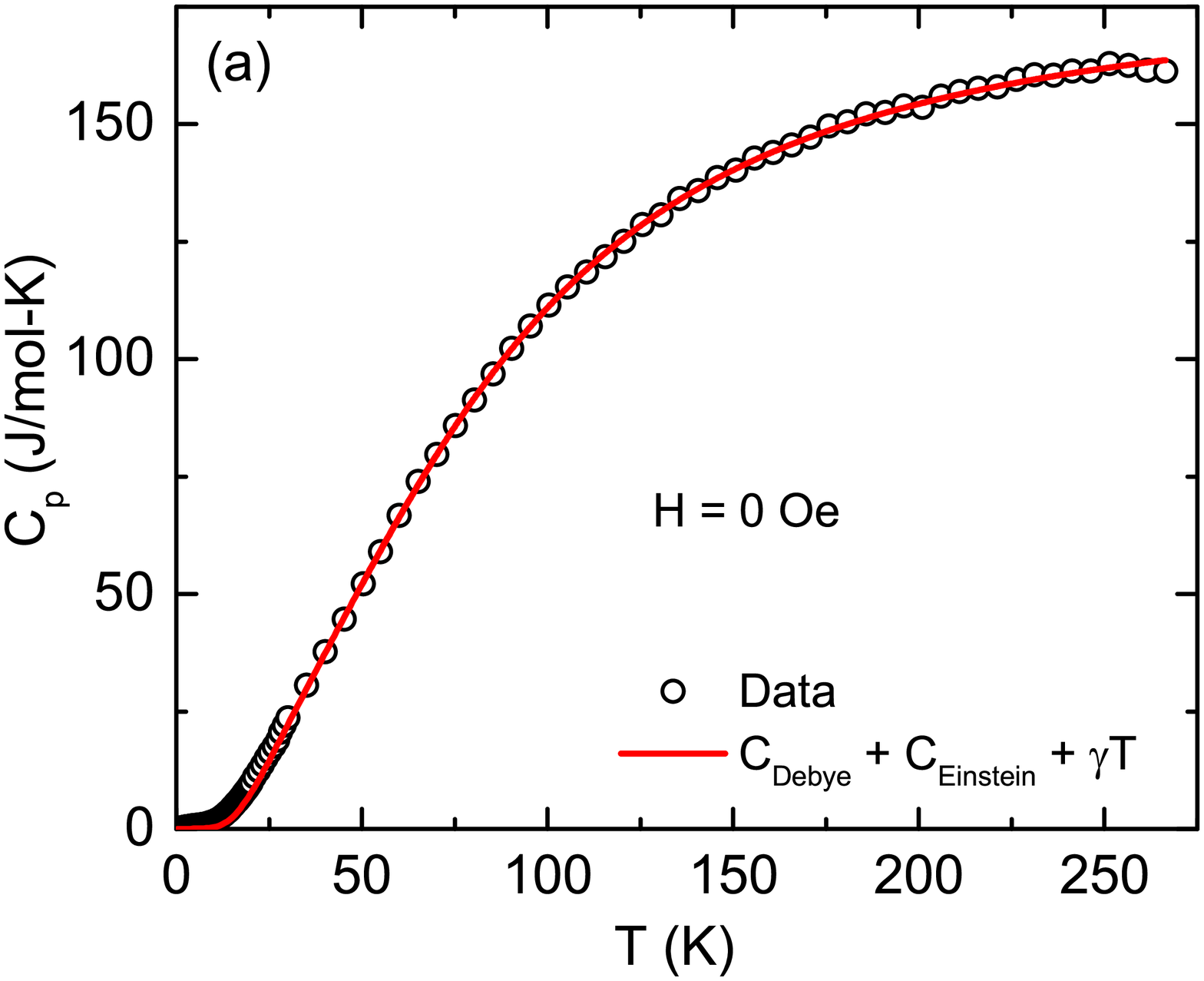}
\includegraphics[width=3.0in]{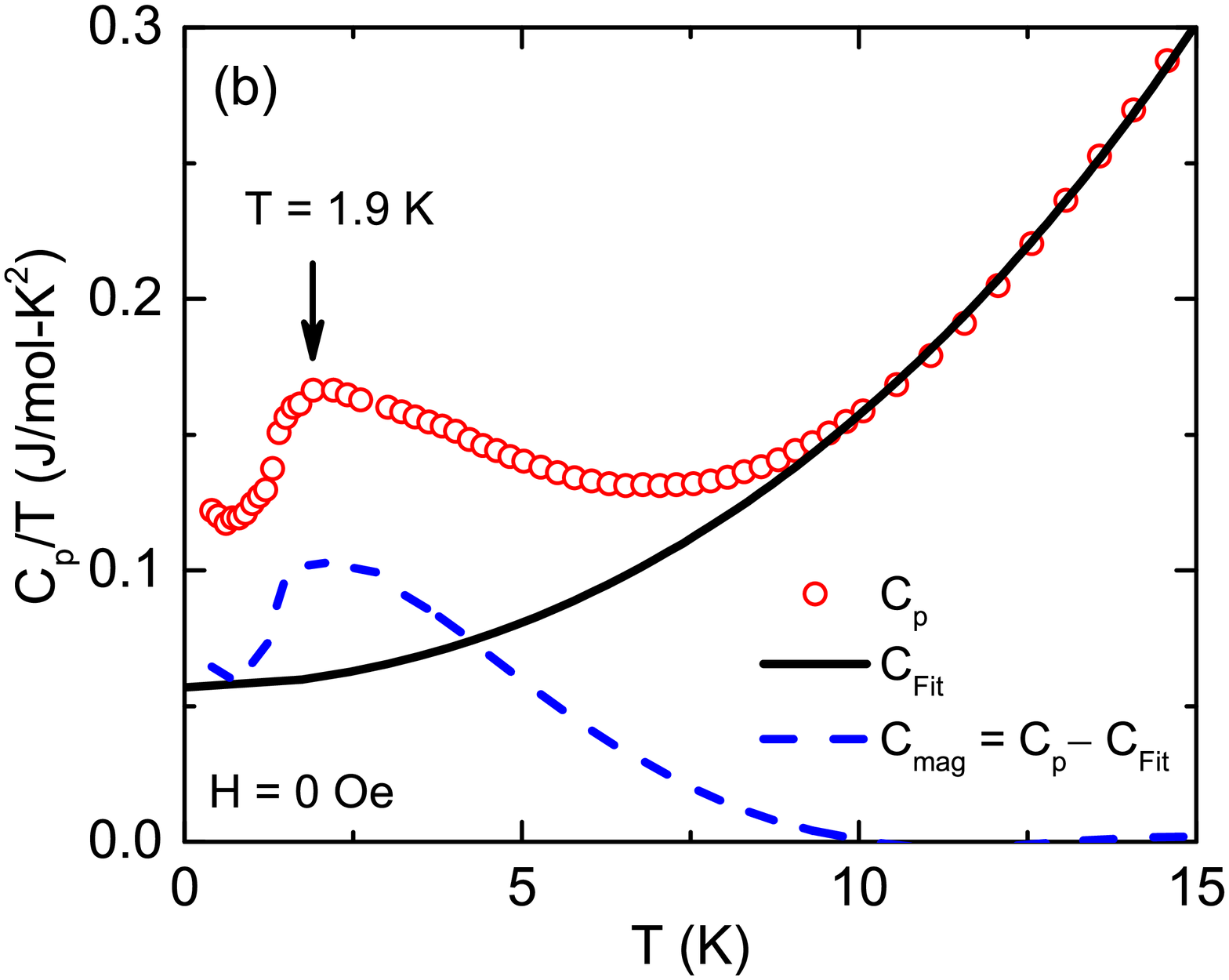}
\includegraphics[width=3.0in]{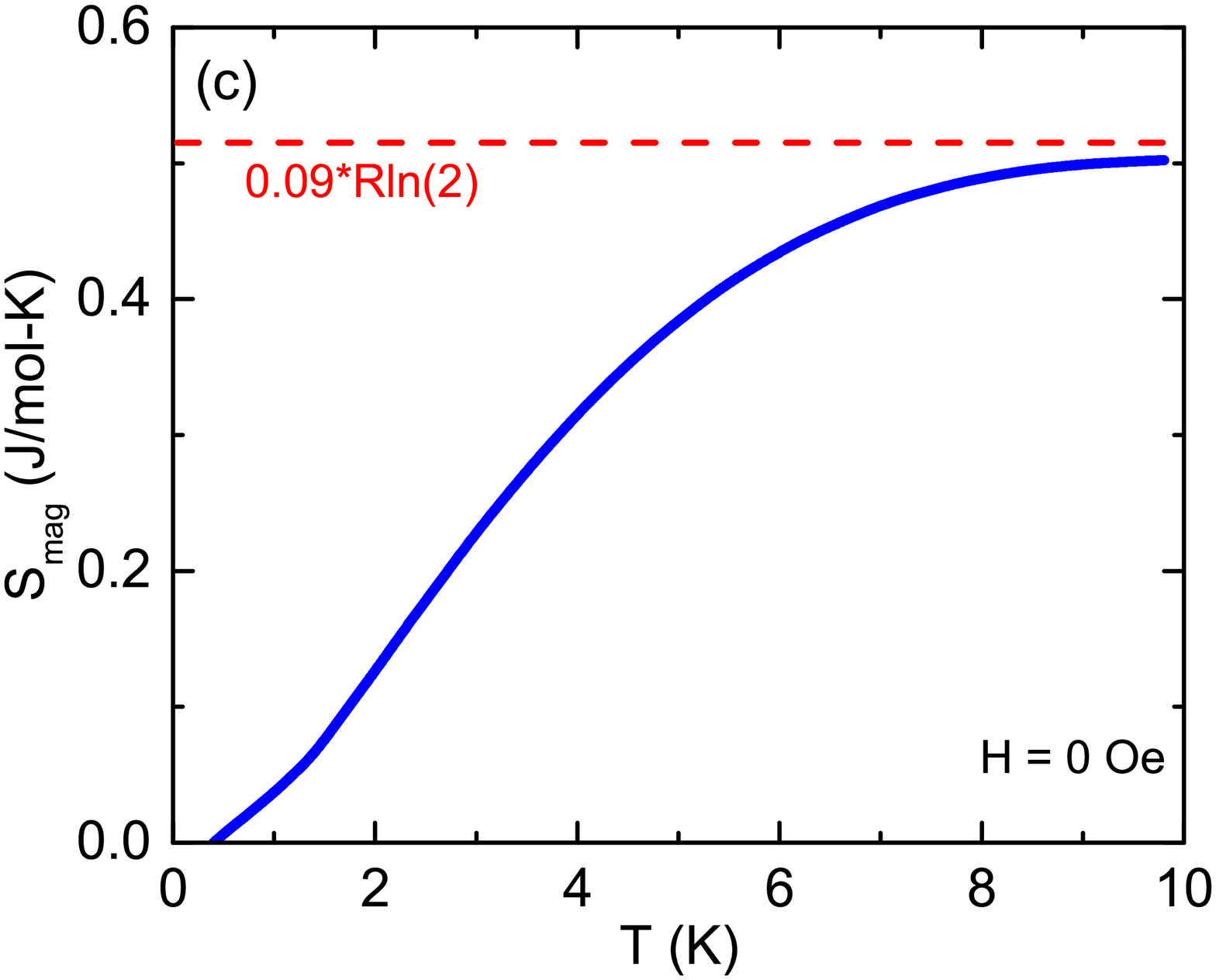}
\caption{(a) Temperature dependence of the molar heat capacity $C_{\rm p}$. The solid curve is the fit including both the Debye model as well as a single frequency Einstein oscillator, as described in the text. (b) Temperature dependence of $C_{\rm p}(T)/T$ for temperatures $0.45~K\leq T \leq 15~K$. Solid line is the fit $C_{\rm p}/T~ = ~\gamma + \beta T^2$, while the dashed curve is the difference of the experimental $C_{\rm p}/T$ and the fitted values. (c) Temperature dependence of the entropy $S_{\rm mag}(T) = \int_{0}^{T}\frac{\Delta C}{T}dT$ extracted from the $C_{\rm p}(T)$ data for $T\leq 10$~K\@. Horizontal dashed line represents 9$\%$ of the magnetic entropy $R{\rm ln}2$ contained in a spin $S = 1/2$ system.}
\label{fig:HC-2}
\end{figure}

The temperature dependence of the specific heat $C_{\rm p}(T)$ for $0.45~K\leq T \leq 266~K$ is presented in Fig.~\ref{fig:HC-2}(a). The value of $C_{\rm p}$ at 266 K is 161(1)~J/mol~K, which is $\sim 8\%$ smaller than the Dulong-Petit limit of heat capacity at constant volume $C_{\rm V} = 3nR = 174.6$~J/mol~K, where $R$ is the gas constant and $n = 7$ is the number of atoms per formula unit. The most likely explanation for this shortfall is that the Debye temperature $\Theta_{\rm D}$ of \TMB\ is significantly larger than 266~K\@. Using the Debye model to fit $C_{p}$(T) data was only partially successful, and it was necessary to add an Einstein term with a fraction $(1-u) = 0.27(7)$ of the total $C_{\rm p}(T)$ to fit the data (Table~\ref{Table:PhysicalProperties}), where $u$ is the fractional contribution of the Debye term. With the above addition, we were able to satisfactorily fit the $C_{\rm p}(T)$ data [Fig.~\ref{fig:HC-2}(a)] with the fitting parameters $\Theta_{\rm D,~HT} = 381(15)$~K, the Debye temperature equivalent of the Einstein mode $\Theta_{\rm E} = 110(22)$~K, fraction of the Debye contribution $u = 0.73(7)$ and $\gamma_{\rm HT}$ = 7(2)~mJ/mol~K$^2$. As expected, the estimated value of $\Theta_{\rm D,~HT}$ is significantly higher than room temperature.

The magnetic contributions to $C_{\rm p}(T)$ are most evident below $\simeq 15$~K, where the high temperature fit discussed above shows a deviation from the measured data [Fig.~\ref{fig:HC-2}(b)]. It is best isolated  by comparing $C_{\rm p}/T$ to the low temperature phonon contribution that is estimated by fitting $C_{\rm p}/T =  \gamma + \beta T^{2}$ over the temperature range $10~K \lesssim T \lesssim 18~K$. This yields the Sommerfeld constant $\gamma = 57(5)$~mJ/mol~K$^{2}$ and $\beta = 0.94(6)$~mJ/mol~K$^4$ (Table~\ref{Table:PhysicalProperties}). The value of the low temperature $\gamma$ is nearly an order of magnitude larger than the value $\gamma_{HT}$ that is inferred at high temperature, suggesting the gradual onset of a correlated electron state with decreasing temperature in \TMB. The value of $\Theta_{\rm D,~LT}$ deduced from the fitted value of $\beta$ is 355(2)~K, in reasonable agreement with the value $\Theta_{\rm D,~HT}$.

At the lowest temperatures, a broad peak centered at $\simeq 1.9~K$ emerges in $C_{\rm p}/T$ [Fig.~\ref{fig:HC-2}(b)]. It is possible that the peak corresponds to the onset of a magnetic order, although the breadth of the peak suggests that the correlations associated with this putative order would be weak and short-ranged. We have plotted in Fig.~\ref{fig:HC-2}(c) the entropy $S_{\rm mag}$ associated with the magnetic part of the specific heat $C_{\rm mag}$, which is obtained by subtracting the low temperature phonon contribution $\beta T^{3}$ and electronic contribution $\gamma T$ from the measured specific heat, $\it{i.e.}$ $C_{\rm mag} = C_{\rm p} - \beta T^{3} - \gamma T$. The $S_{\rm mag}$ saturates to 0.09$R$ln2 around 10 K\@. If the peak in the specific heat $C_{\rm p}(T)$ at 1.9 K is associated with magnetic ordering, then it is clear that the ordered moment is significantly smaller than the spin 1/2 moment of  1.73~$\mu_{B}$ whose entropy would be $R$ln2.

\subsection{Electronic Structure Calculations}
\begin{figure*}
\centering
\includegraphics[width=6in]{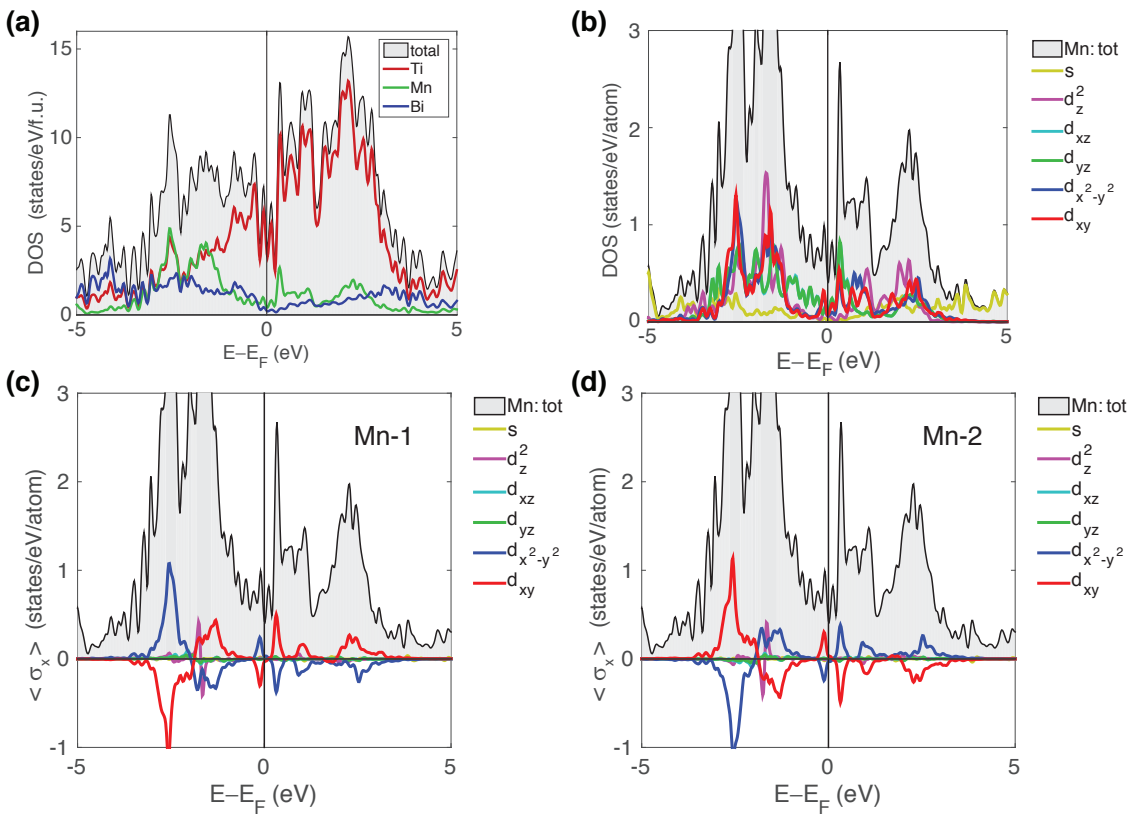}
\caption{(a) Atom decomposed density of states (DOS) of Ti, Mn, and Bi as well as the total DOS, which clearly demonstrates that Ti contributes most of the DOS at the Fermi level. (b) Orbitally differentiated Mn states that are mostly centered around $\pm 2.5$~eV. The breadth of these states is comparable to their energies, indicating their pronounced itinerant character. To further understand the magnetic ordering, (c) and (d) show the the spin- and orbital-resolved DOS for two intra-chain neighboring Mn atom represented as Mn1 and Mn2, respectively, where $\sigma_x$ indicates spin moment of the orbitals along the $x$ direction. Only the  $d_{xy}$ and $d_{x^{2}-y^{2}}$ orbitals, lying in the $ab$-plane  have significant spin polarization. Antiferromagnetic order is predicted, since the spin polarization of these orbitals is opposite on neighboring Mn1 and Mn2 atoms.}
\label{fig:dos}
\end{figure*}

The total and atom-decomposed electronic density of states (DOS) of \TMB\ are shown in Fig.~\ref{fig:dos}(a). The compound is definitively metallic, having a sizable density of states ${\cal D}_{\rm band}(E_{\rm F})$ = 7.4(1)~states/(eV f.u.) for both spin directions at the Fermi energy $E_{\rm F}$. To understand the contribution of the specific atoms to the DOS, we transformed the Kohn-Sham eigenstates of electronic structure calculations (ESC) into a set of highly localized quasiatomic orbitals. The eigenvalues of the Kohn-Sham wavefunctions were transformed into a first-principles tight-binding Hamiltonian \cite{marzari2012,qian2008} which was subsequently used for computing the results. Figure~\ref{fig:dos}(a) demonstrates that the Ti atoms contribute most of the DOS close to $E_{\rm F}$, while the Mn and Bi atoms have much smaller shares. Within the accuracy of the ESC, none of the Mn $d$-orbitals makes a substantial contribution to the DOS at the $E_{\rm F}$, which is instead dominated by the Ti states that appear to hybridize much more strongly with each other than with the Mn states.

Our measurements of the electrical resistivity and specific heat both suggest that the conduction electrons have significant electronic correlations at low temperatures. To further test this point, we estimated the DOS at $E_{\rm F}$ from the experimental value of $\gamma$,

\begin{equation}
{\cal D}_\gamma(E_{\rm F})  = \frac{{\pi^2}k_{\rm B}^2}{3\gamma},
\end{equation}

that includes the many-body enhancement effects such as electron-electron and electron-phonon interactions. From the estimated $\gamma$ of \TMB\ (Table~\ref{Table:PhysicalProperties}), we calculate ${\cal D}_\gamma(E_{\rm F}) = 24(3)$~states/(eV f.u.) for both spin directions. The estimated ${\cal D}_\gamma(E_{\rm F})$ is significantly larger than the ${\cal D}_{\rm band}(E_{\rm F}) = 7.4(1)$~states/(eV f.u.) obtained from the ESC, confirming the presence of many-body enhancement effects in this material. We get ${\cal D}_\gamma(E_{\rm F})/{\cal D}_{\rm band}(E_{\rm F}) = \gamma/\gamma_{\rm band} = (1+\lambda_{\rm el-ph})m^{*}/m_{\rm band} = 3.2(5)$, where $\gamma_{\rm band}$ and $m_{\rm band}$ are band theory values of the Sommerfeld coefficient and conduction carrier effective mass and $\lambda_{\rm e-ph}$ is the electron phonon coupling constant. A factor of three enhancement in electron mass is in line with values computed and experimentally deduced for iron pnictides and chalcogenides~\cite{yin2011} and also their Mn analogs~\cite{mcnally2014,Sangeetha-2016,mcnally2015}.

The orbitally differentiated density of Mn states [Fig.~\ref{fig:dos}(b)] is centered at $\simeq 2.5$~eV from the $E_{\rm F}$, reflecting the strength of the Coulomb and Hunds interactions that stabilize magnetic moments within the correlated Mn-based bands. Hybridization produces a substantial breadth to these states, extending from about $-3$ to $+3$~eV, indicating that charge and spin excitations are strong in \TMB. Apparently, the Mn states are not fully localized and lack a stable valence, which is a defining characteristic of metals with strong correlations and of insulators.

Calculation of the spin-polarized density of states reveals that the \TMB\ may be magnetically ordered. The spin- and orbital-resolved DOS for the two Mn atoms are plotted in Figs.~\ref{fig:dos}(c) and (d). A hidden magnetic ordering is revealed, in that only the $d_{x^2-y^2}$ and $d_{xy}$ states are able to sustain spin polarization. Comparing the same $d$-orbitals in Fig. 4(c) and (d), it is seen that the sign of the spin magnetic moment of the two intra-chain neighboring Mn atoms, represented as Mn-1 and Mn-2, is reversed. Although the total spin magnetic moment on each Mn is small compared to the overall Mn moment, the ESC predicts that AFM ordering will be established on the $d_{x^2-y^2}$ and $d_{xy}$ orbitals.

\subsection{Magnetic Properties}
\label{SSection:Magnetism}
\begin{figure*}
\includegraphics[width=6.6in]{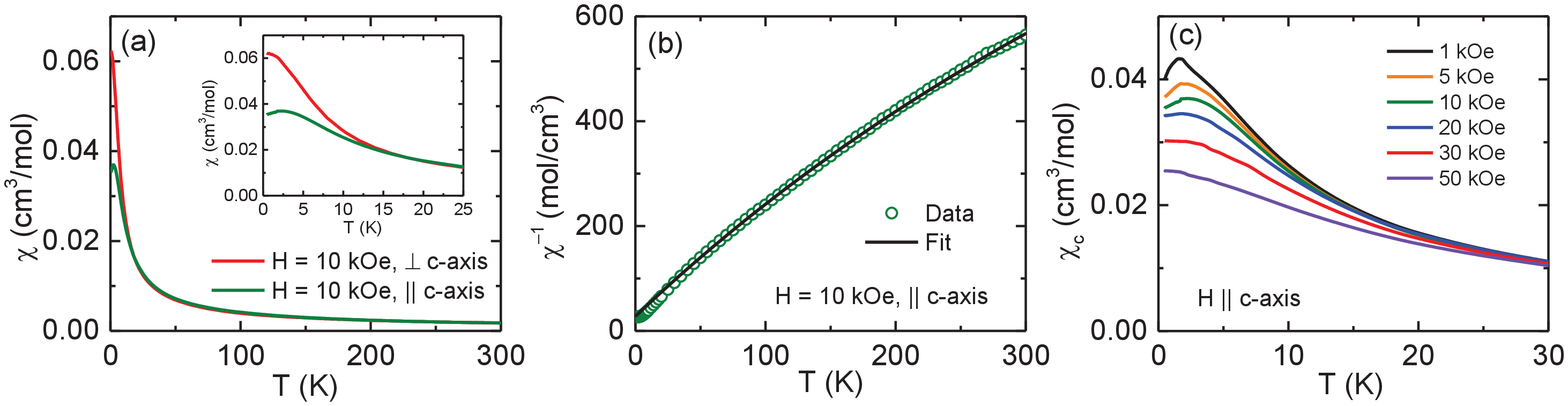}
\caption{(a) Temperature dependence of the magnetic susceptibility $\chi = M/H$ of \TMB. The red data have a 10~kOe field applied perpendicular to the $c$-axis, while the green data have the same field applied along the $c$-axis. Inset: an expanded view of $\chi(T)$ at low temperatures highlighting its anisotropy. (b) Temperature dependence of $\chi^{-1}$ at 10~kOe field applied parallel to the $c$-axis. Solid curve is the Curie-Weiss fit as described in the text. (c) Temperature dependencies of $\chi_{c}(T)$ measured at six different fields between 1 and 50~kOe, as indicated. All fields were applied parallel to the $c$-axis.}
\label{fig:Mag-1}
\end{figure*}

Temperature dependence of the magnetic susceptibility $\chi(T)$ of \TMB\ measured with a 10~kOe magnetic field oriented both perpendicular and parallel to the $c$-axis is plotted in Fig.~\ref{fig:Mag-1}(a). In both field orientations, broad maxima are observed that are centered near 1.9~K\@. Above $\simeq 15~K$, $\chi_{ab}(T)$ and $\chi_{c}(T)$ data are nearly identical. The inverse of the susceptibility $\chi^{-1}(T)$ data are well fitted by the Curie-Weiss expression augmented by a temperature independent contribution $\chi_{0}$,

\begin{equation}
\chi(T) = \frac{C}{T-\theta_{\rm p}} + \chi_{0},
\label{eq:InvSus}
\end{equation}

where $C$ is the Curie constant and $\theta_{\rm p}$ is the \mbox{paramagnetic} Weiss temperature. The fitted values of the parameters are as follows: $C = 0.41(2)$~cm$^{3}$~K/mol, $\theta_{\rm p} = -9(3)$~K and $\chi_{0} = 4.8(5)\times10^{-4}$~cm$^{3}$/mol (Table~\ref{Table:PhysicalProperties}). The estimated value of the effective paramagnetic moment $\mu_{\rm eff} = \sqrt{8C}$ (assuming $g = 2$) is 1.79(4)~$\mu_{\rm B}$/Mn,  which is close to the 1.73~$\mu_{\rm B}$ expected for a spin $S = 1/2$ system. This suggests that the fluctuating moments are associated with the Mn$^{2+}$ ions, which are likely in a low spin configuration. The negative value of $\theta_{\rm p}$ establishes that the dominant interaction in this material is of AFM nature. Within the Heisenberg model of crystallographically equivalent localized spins, the exchange constant(s) can be estimated using $\theta_{\rm p} = -\frac{S(S+1)}{3k_{\rm B}}\sum_{j}{J_{ij}}$ \cite{Johnston-2012}. Assuming that the dominant exchange $J_{\rm c}$ couples magnetic moments along the chains where the distance between Mn moments is smallest, we estimate $\theta_{\rm p} = - J_{\rm c}/2k_{\rm B}$ for $S = 1/2$, which leads to $J_{\rm c} = 18(6)$~K\@. 

The Pauli susceptibility of Ti$_4$MnBi$_2$ estimated using $\chi_{\rm Pauli} = \mu_{\rm B}^{2} {\cal D}(E_{\rm F}) = 7.8(9)\times10^{-4}~{\rm cm^3/mol}$ for ${\cal D}_{\gamma}(E_{\rm F}) = 24(3)~{\rm state/eV~f.u.}$ is similar to the value of $\chi_{0}$ (Table~\ref{Table:PhysicalProperties}), suggesting the same origin for both.  The fact that both the ${\cal D}_{\chi_0}(E_{\rm F})$ = 14.8 states/eV~f.u. and the ${\cal D}_{\gamma}(E_{\rm F})$ show a substantial enhancement over the ${\cal D}_{\rm band}(E_{\rm F})$ further suggests presence of sizable electron correlation effects in \TMB. The inset of Fig.~\ref{fig:Mag-1}(a) shows that at low temperatures, the $\chi_{ab}$ measured with the field in the basal plane ($H\perp c$) has a much stronger temperature dependence compared to the $\chi_{c}$ when the field is along the $c$-axis ($H\parallel c$). The $\chi_{0}$ is subtracted from the measured anisotropic susceptibilities to isolate their temperature dependent parts, which demonstrates that the anisotropy approaches $\sim 2$ at 1.9~K\@.  The measurements with $H\parallel c$ are reported for different fields between 1 and 50~kOe in Fig.~\ref{fig:Mag-1}(c). An abrupt slope change in $\chi_{c}(T)$ measured with $H = 1$~kOe indicates the onset of AFM order at $T_{\rm N} = 1.9$~K. Increasing magnetic fields broaden and weaken the peak in $\chi_{c}(T)$, suggesting that the magnetic correlations become progressively short ranged as field destabilizes AFM order.

\begin{figure}
\includegraphics[width=3.3in]{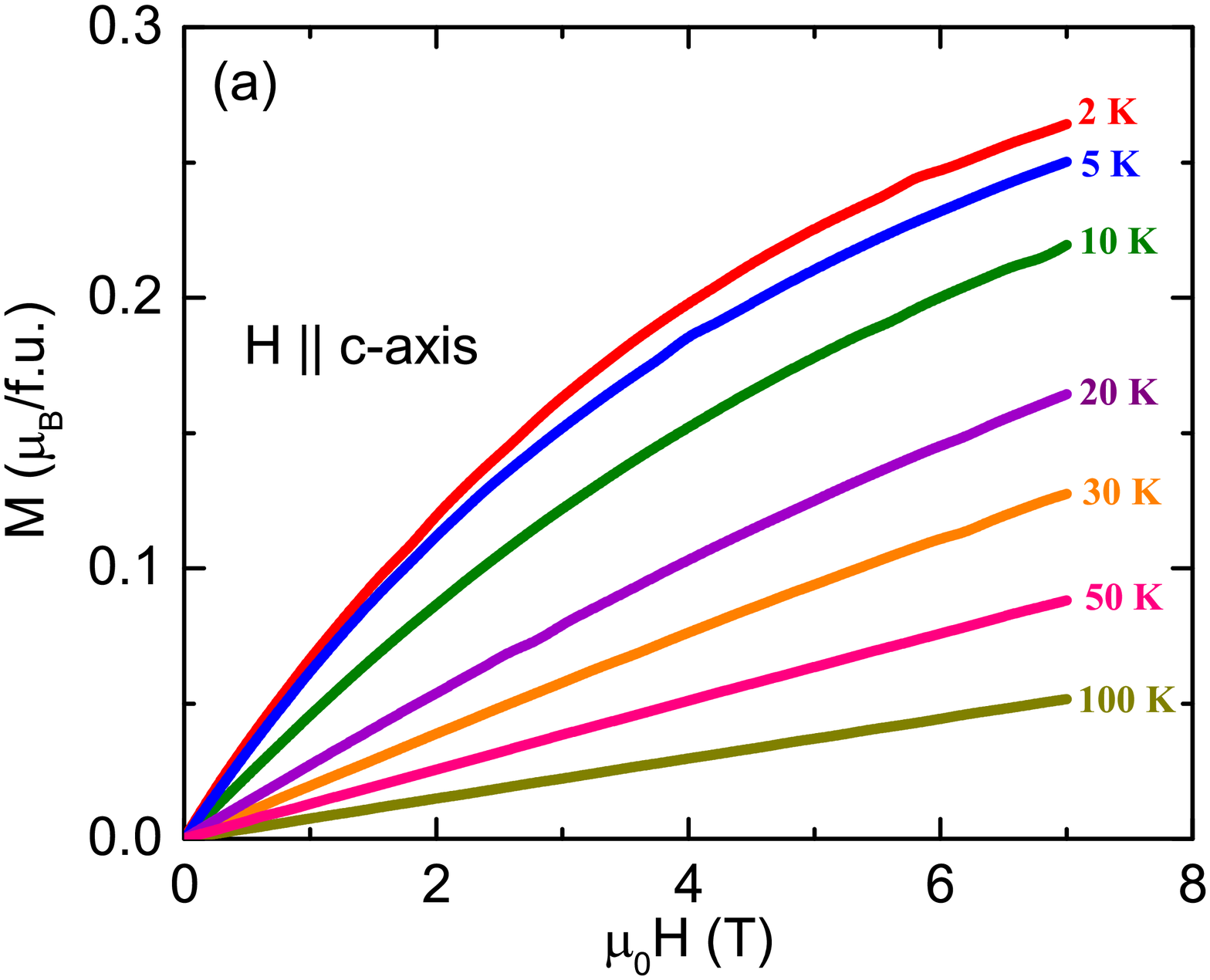}
\includegraphics[width=3.3in]{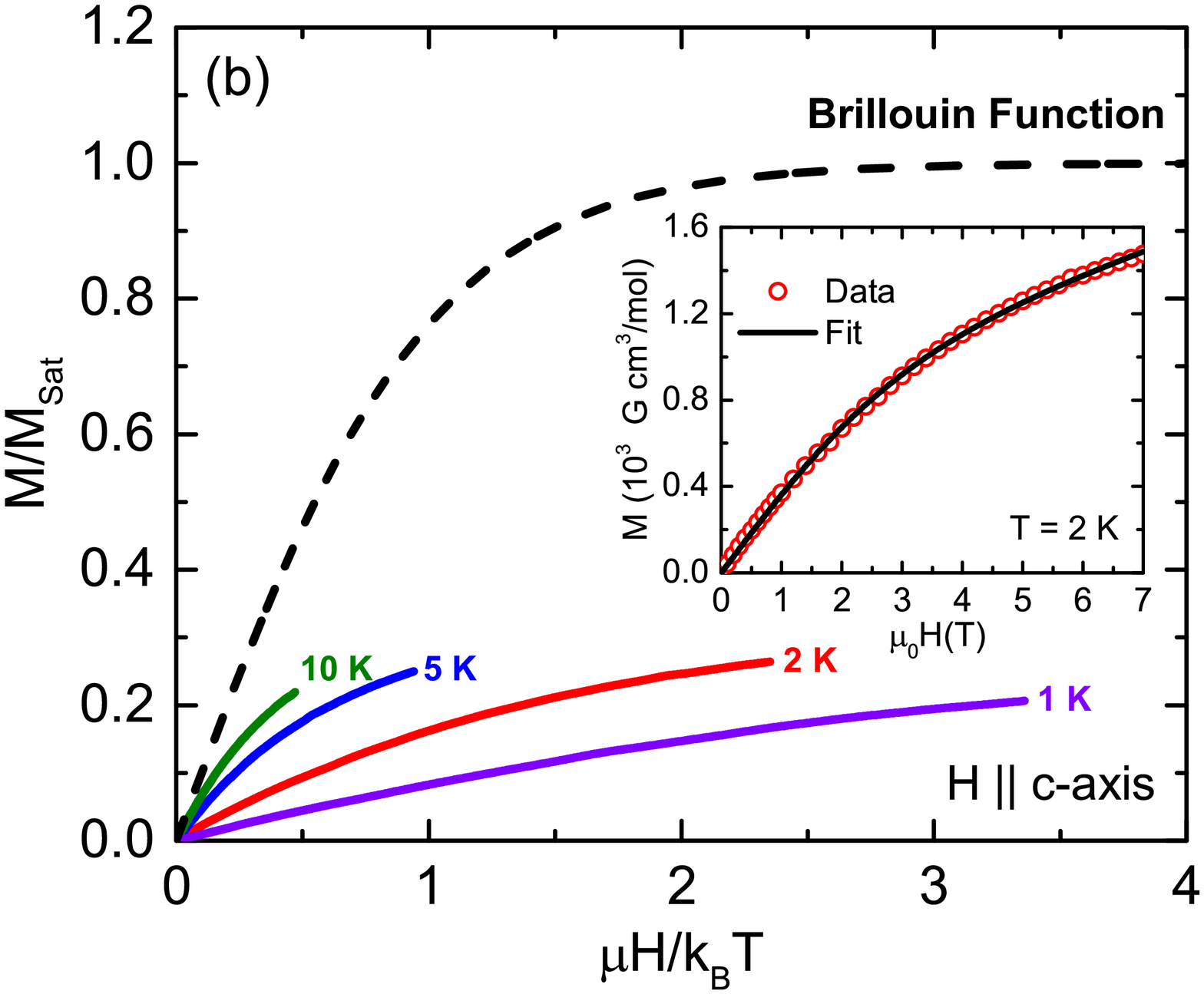}
\caption{(a) Isothermal magnetization $M(H)$ measured with the field $H\parallel~c$-axis at seven different temperatures, as indicated. (b) Normalized $M/M_{\rm Sat}$ plotted as a function of $\mu H/k_{\rm B}T$ at low temperatures, as indicated, where $M_{\rm Sat} = gS\mu_{\rm B}$. Simulated Brillouin function with $S = 1/2$ and $g = 2$ is shown by the dashed line. Inset: the fit of the $M(H)$ data at 2~K as described in the text.}
\label{fig:Mag-2}
\end{figure}

The field dependence of the magnetization $M$ provides further information about the magnetic correlations and moment in \TMB. Isotherms of the magnetization $M(H)$ are shown in Fig.~\ref{fig:Mag-2}(a) for seven different temperatures between from 2 to 100~K\@. The $M(H)$ plots are nonlinear and the nonlinearity becomes more pronounced as the temperature decreases. Isothermal magnetization normalized to the saturation magnetization $M/M_{\rm sat}$ is plotted as a function of the  scaled magnetic energy $\mu H/k_{\rm B}T$ (where $M_{\rm sat} = gS\mu_{\rm B} = 1~\mu_{\rm B}$ considering $S = 1/2$ and $g = 2$) at four different temperatures between 1 to 10~K [Fig.~\ref{fig:Mag-2}(b)]. For comparison, we have also plotted the simulated Brillouin function (BF) for $S = 1/2$, which represents the behavior of localized spins in the paramagnetic regime. It is evident from the plot that the magnitude of $M/M_{\rm sat}$ at any measured value of $\mu H/k_{\rm B}T$ is significantly reduced compared to the expectations of the BF [Fig.~\ref{fig:Mag-2}(b)]. Apparently, there is a non-saturating contribution to the magnetization data at the low temperatures that results in the positive gradient in the $M(H)$ curves even at the high fields of our measurement. A plausible reason for the observed reduction is the emergence of AFM correlations that also show their signature in the $C_{\rm p}(T)$ and in the $\chi(T)$ measurements at the low temperatures. To investigate this scenario, we fitted the $M(H)$ data taken in close proximity ($T = 2$~K) to the magnetic transition temperature ($T_{\rm N} = 1.9$~K) using the expression,

\begin{subequations}
\label{eq:M-Expression}
\begin{equation}
M(T, H) = fM_{\rm sat}B_{S}\left[\frac{g\mu_{\rm B}H}{k_{\rm B}T}\right] + \chi_{\rm AF} H,
\label{eq:M-Fit}
\end{equation}

where $f$ is a prefactor to the BF and $\chi_{\rm AF}$ represents the susceptibility of the antiferromagnetically coupled spins \cite{WeissTemp-BF}. The BF can be described as \cite{Goetsch-2012}, 

\begin{equation}
B_{S}(x) = \frac{1}{2S}\left\{(2S+1){\rm coth}\left[(2S+1)\frac{x}{2}\right] - {\rm coth}\left(\frac{x}{2}\right)\right\},
\label{eq:BF}
\end{equation}
\end{subequations}

where $x = g\mu_{\rm B}H/k_{\rm B}T$ and $M_{\rm sat} = N_{\rm A}gS\mu_{\rm B}$. The $M(H)$ data of \TMB~ measured at $T = 2$~K were fitted satisfactorily [inset, Fig.~\ref{fig:Mag-2}(b)] to this expression by varying only two parameters $f$ and $\chi_{\rm AF}$ and keeping $g = 2$ and $S = 1/2$. As anticipated from the Fig.~\ref{fig:Mag-2}(b), the weighting factor $f$ is much smaller than one [$f=0.146(2)$] and $\chi_{\rm AF} = 9.8(2)x10^{-3}$~cm$^{3}$/mol.  

This analysis suggests an intermediate degree of localization in \TMB. The appearance of the term in $M(H)$ that can be described by the BF attests to the presence of reasonably well defined Mn moments. Unlike an insulator where $f = 1$ and each moment is localized by correlations into a well defined and long-lived valence state, the observation in \TMB\ that $f \ll 1$ is consistent with the breadth in energy of the Mn states [Fig.~\ref{fig:dos}(b)] where the valence is poorly defined and the Mn $d$-electrons have a pronounced itinerant character due to hybridization effects. While it is a matter for future work, it is possible that quantum fluctuations associated with either the one-dimensionality of \TMB\ or alternatively to proximity to an electronic localization instability that produces a magnetic moment from a correlated band could contribute additional breadth to the Mn states at the lowest temperatures. One may wonder that the second term $\chi_{\rm AF}H$ could possibly be associated with the Pauli susceptibility of the conduction carriers. However, $\chi_{\rm AF}$ is nearly an order of magnitude larger than the value of $\chi_{\rm Pauli}$ (Table~\ref{Table:PhysicalProperties}), suggesting that $\chi_{\rm AF}$ likely reflects the presence of AFM correlations or even long ranged order that affects the partially localized Mn moments.

\begin{figure*}
\includegraphics[width=6.6in]{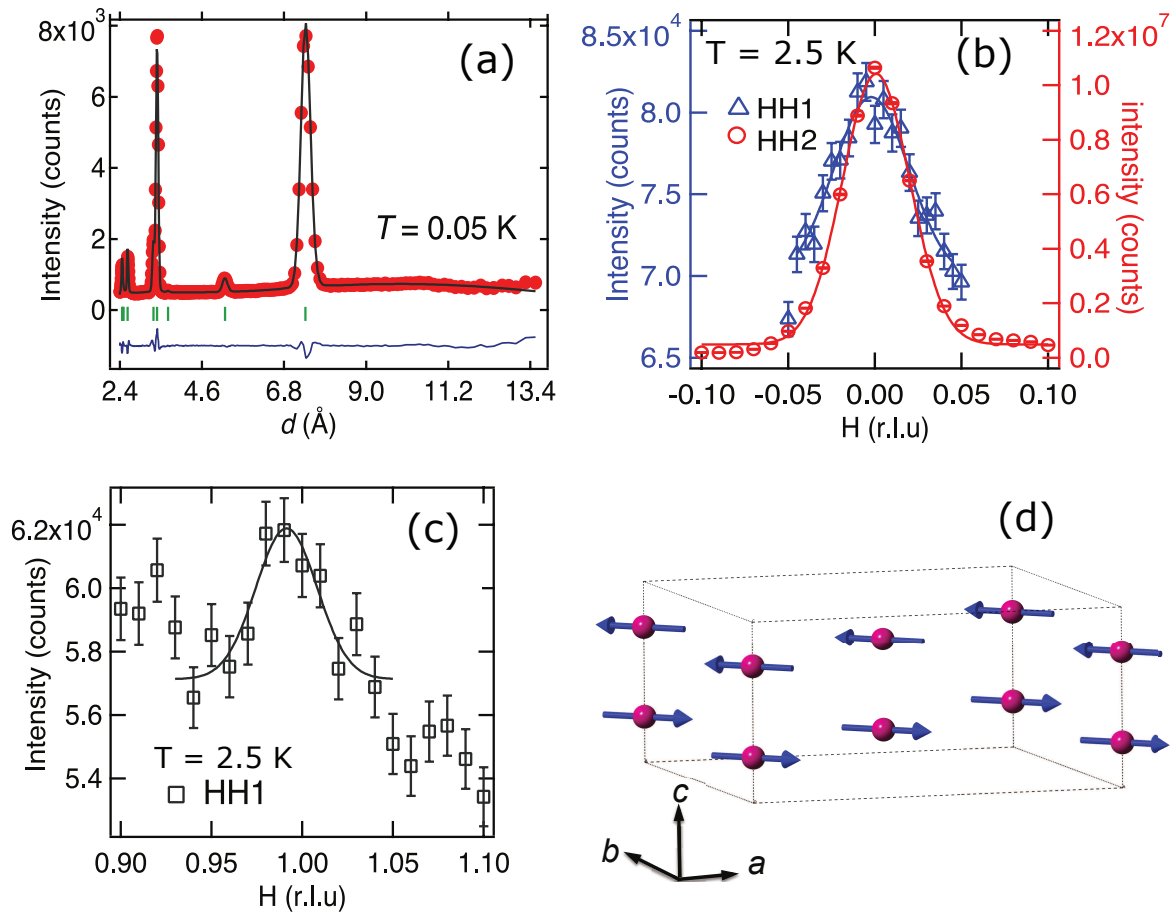}
\caption{(a) Rietveld refinement of the neutron powder diffraction data of \TMB\ measured at 0.05~K\@. Experimental data points are shown by solid symbols, the black curve through them is a Rietveld analysis fit, green vertical bars indicate nuclear Bragg reflections (space group: $I{\rm 4}/mcm$) and blue curve is the difference between the experimental and the calculated profiles. (b) Single crystal neutron diffraction data obtained at 2.5 K showing the existence of a weak magnetic peak at (001). The nuclear reflection (002) is also shown for comparison. (c) Another weak magnetic peak (111) is observed in the single crystal neutron diffraction experiments. Both (001) and (111) reflections are forbidden by the crystal structure of \TMB. (d) Magnetic structure for \TMB\ determined from representation analysis (Supplementary Materials), showing Mn moments with antiferromagnetic alignment along the $c$-axis and ferromagnetic alignment within the $ab$-plane.}
\label{fig:ND}
\end{figure*}

Neutron diffraction measurements have been used to search for the presence of magnetic order in \TMB. Figure~\ref{fig:ND}(a) shows the neutron powder pattern obtained at $0.05$~K, where the data were fit using Rietveld refinement. The estimated agreement factor of the fit is $R_{\rm wp} = 4.85\%$, and the diffraction peaks can all be indexed as the nuclear peaks of the crystal structure. Furthermore, there is no indication of additional peaks or temperature dependencies that might suggest the onset of magnetic order between 6.5 and $0.05$~K (Fig.~S2, Supplementary Materials). Neutron diffraction scans were carried out at 2.5~K along selected directions of reciprocal space, using an aligned collection of single crystals. Several weak and broad peaks were observed near the $001$ [Fig.~\ref{fig:ND}(b)] and $111$ [Fig.~\ref{fig:ND}(c)] positions. Since these data were obtained above the apparent AFM ordering temperature of 1.9~K deduced from the maxima in $C_{\rm p}/T$ and $\chi(T)$, these moments are correlated over smaller length scales, poised to turn into true magnetic diffraction peaks when the temperature drops below $T_{\rm N}$. Since these reflections are forbidden in the $I4/mcm$ space group of \TMB, it is concluded that the propagation wave vector of the implied magnetic order is either (1,1,1) or (0,0,1). A symmetry analysis was performed using the program SARAh~\cite{wills2000} (see the Supplementary Materials). Of the eight irreducible representations appropriate for \TMB, only one can reproduce the relative intensities of the (001) and (111) diffraction peaks. 

The basis vectors imply that the ordered Mn moment lies in the $\it{ab}$-plane, as do the lobes of the associated $d_{xy}$ and $d_{x^2-y^2}$ orbitals. As shown in Fig.~\ref{fig:ND}(d), this magnetic structure has Mn moments that are perpendicular to the $c$-axis. The Mn moments are antiferromagnetically aligned along the $c$-axis but ferromagnetically correlated within the $ab$-plane. Thus, the dominant feature of the magnetic structure of \TMB\ is the presence of AFM order that doubles the chemical unit cell along the $c$-axis, with the FM order between the chains. The ratio of the structure factors obtained for the (001) and (002) diffraction peaks provides the normalization for the AFM peaks and in this way we obtained an ordered moment of $\simeq~0.05~\mu_{\rm B}$/Mn. The small ordered moment explains its apparent absence in the powder diffraction experiment, which would typically only be sensitive to ordered moments $\mu \sim 0.1~\mu_{\rm B}$. It is significant that the ordered moment $0.05~\mu_{\rm B}$/Mn is only $3\%$ of the 1.79 $\mu_{B}$/Mn Curie-Weiss moment, suggesting that the magnetic fluctuations are very strong in \TMB, to the point that they are almost overwhelming magnetic order.

\section{Discussion}

The specific heat [Fig.~\ref{fig:HC-2}(b)] and the magnetic susceptibility [Fig.~\ref{fig:Mag-1}(a)] measurements have broad peaks centered near 1.9~K that provide corroborating evidence for the magnetic order that has been found in the neutron diffraction measurements. The entropy contained in the specific heat peak [Fig.~\ref{fig:HC-2}(c)] is about $9\%$ of $R$ln2/Mn, providing a measure of the staggered moment that is possibly more accurate than that found in the single crystal neutron diffraction measurements, which were obtained for $T~\simeq~T_{\rm N}$ where the staggered moment would not yet be fully developed as at $T \ll T_{\rm N}$. The qualitative conclusion is unaffected, that the ordered moment is much smaller than the Curie-Weiss moment of $1.79~\mu_{\rm B}$/Mn. The breadth of the peaks in the single crystal neutron diffraction measurements as well as in the temperature dependencies of the specific heat and magnetic susceptibility are suggestive that a true long-ranged order does not occur in \TMB, rather the correlations remain finite ranged due to strong fluctuations. These fluctuations could potentially be the result of magnetic frustration that originates with the low dimensionality implied by the chain-like crystal structure of \TMB. However, the ESC raise the alternative possibility that \TMB\ could be naturally tuned to an instability of the Fermi surface itself, leading to the emergence of a tiny Mn moment from a correlated band of Mn states.

Resistivity and heat capacity measurements establish that \TMB\ is inherently metallic and this conclusion is confirmed by the ESC. The calculations find that the DOS at the $E_{\rm F}$ is dominated by Ti states, which are substantially hybridized. The comparison to experiment indicates that these Ti-based states have moderate electronic correlations. The DOS at the $E_{\rm F}$ determined from the Sommerfeld coefficient $D_{\gamma}(E_{\rm F}) = 24$~states/(eV f.u.) is enhanced by a factor of 3.2 relative to that determined from the ESC $D_{\rm band}(E_{\rm F}) = 7.4$~states/(eV f.u.), suggesting substantial electronic correlations that are not captured by the ESC. The Kadowaki-Woods ratio $R_{\rm KW}$ = 0.34$~\mu\Omega$~cm~mol$^2$J$^{-2}$K$^2$ is much smaller than the value of 10~$\mu\Omega$~cm~mol$^2$J$^{-2}$K$^2$ that is typical of the highly correlated heavy electron compounds, but is similar to the value of 0.3 $\mu\Omega$cm~mol$^2$J$^{-2}$K$^2$ that is reported for Pd and other transition metal based compounds \cite{jacko2009}. The Sommerfeld-Wilson ratio $R_{\rm W}$ is given by $R_{\rm W}$ = $\pi^{2}/3$($k_{\rm B}^{2}$/$\mu_{\rm B}^{2}) (\chi_{0}/\gamma)$, where $k_{\rm B}$ is Boltzmann's constant and $\mu_{\rm B}$ is the Bohr magneton. In an uncorrelated Fermi liquid $R_{\rm W} = 1$, but $R_{\rm W}$ is greater than 1 if FM correlations are present. Using the values $\gamma$ = 57 mJ/mol K$^{2}$ and $\chi_{0}$ = 4.8$\times$ 10$^{-4}$ cm$^{3}$/mol, we find that $R_{\rm W}$ = 0.6(2), ruling out appreciable FM correlations in \TMB.

A range of magnetic characters is possible for Mn in metals that reflects the interplay of local Hund's and Coulomb interactions, which tend to localize magnetic moments on the Mn sites, with the hybridization of the Mn states, reflected in an effective bandwidth. When the onsite interactions are very strong and the hybridization weak, robust insulators like LaMnPO~\cite{mcnally2014}, CaMn$_{2}$Sb$_{2}$~\cite{simonson2012} and BaMn$_{2}$As$_{2}$ ~\cite{mcnally2015, Ann-2009, Singh-2009} occur where nearly full Hund's rule moments are found. When the hybridization is strong as in metallic Mn$X$ ($X$ = P, As, Sb, Bi) \cite{okuda1999, shanavas2014,okabayashi2004}, Rb-doped BaMn$_2$As$_2$ \cite{Pandey-2015}, MnSi \cite{carbone2006}, YMn$_{2}$ \cite{yamaoka2009} and HfMnGa$_{2}$ \cite{marques2011}, residual onsite interactions can induce a Mn moment within the correlated electron bands that can be substantially weaker than the Hund's rule values. Assuming that the bandwidth reflects only Mn-Mn hybridization, a Mn moment emerges from the band for a critical value of hybridization that occurs for Mn-Mn spacings near 2.5~$\AA$, antiferromagnetism and growing Mn moments are found for intermediate values of Mn-Mn spacing between $2.5-2.8~\AA$, while fully localized moments and strong FM character are found for well separated Mn ions with Mn-Mn spacings larger than 2.9~$\AA$ \cite{coey2010}. The nearest Mn-Mn spacing in \TMB\ is 2.5~$\AA$ along the chain direction and in this scheme, \TMB\ might be expected to be on the verge of an electronic transition that would produce a somewhat localized moment, possibly leading to weak AFM order. These expectations are upheld by the ESC, where the similarity of the  Mn state energy and bandwidth attests to a modest level of correlations, with just sufficient localization to provide a Curie-Weiss moment of 1.79~$\mu_{\rm B}$/Mn. The observations of weak AFM order and staggered moments that are much smaller than the Hund's rule moments are suggestive that \TMB\ naturally forms close to the instability that leads to the formation of an AFM quantum critical point (QCP) where moments form and simultaneously order at $T = 0$. The ESC support this finding, where the Mn density of states is found to be spin polarized, supporting a small staggered moment, in qualitative agreement with the value inferred from the entropy $S_{\rm mag}$ associated with AFM order and from the staggered moment found in neutron diffraction measurements. It is interesting to note that the staggered moment found by ESC is associated only with the $d_{x^{2}-y^{2}}$ and $d_{xy}$ orbitals, while the remaining spectral weight associated with the Mn $d$-electrons forms a broad band centered near $-2.5$~eV, with a band width of approximately 3~eV. It is important to understand that there are two Mn-based $d$ orbitals that are excluded from this scenario, and that their susceptibility to magnetic polarization is responsible for the weak AFM order that is found in the neutron diffraction measurements in \TMB. The lobes of these orbitals lie in the plane perpendicular to the Mn-chain axis, in agreement with the magnetic structure deduced from the neutron diffraction measurements. It remains an open question whether the chain-like structure of \TMB\ results in one-dimensionality of the electronic or magnetic subsystems.

The crystal structure of \TMB\ features chains of Mn atoms, with large interchain separations suggesting that 1D physics may be relevant to our understanding of this remarkable material. If so, it is one of only a few 1D systems that are metallic, and it is potentially the first member of a larger and previously unexplored family of 1D systems. It remains a largely open question as to how electronic correlations, familiar to us in 3D systems, affect the lowest energy excitations in 1D. By analogy, it seems likely that the ESC in 1D systems provide essential information regarding the states that are involved in bonding, the stability of magnetic moments and whether there is an insulating gap or not. Similarly, there is a separation in scales between these high energy states where dimensionality plays little role and the fundamental excitations that are found at the lowest energies, where the 1D character is of paramount importance. In most spin chain systems studied so far ~\cite{Hirakawa-1967, Lake-2005, Lee-2012, Pavarini-2008, Kim-1996, Schlappa-2012,mourigal2013}, their insulating character implies that onsite Hund's and Coulomb interactions are very strong compared to hybridization. It follows that charge excitations are profoundly gapped at high energies and temperature, although any spinons or holons present at low energies can be gapped or not depending on the anisotropy of the underlying XXZ Hamiltonian, or the strength of the  interchain coupling~\cite{Giamarchi-2004}. In this case, we expect two different Fermi surfaces for the spinons and holons.

Alternatively, if the bandwidth of states with mobile charge is comparable to the strength of the Coulomb interactions in organic 1D conductors, as is potentially the case for \TMB, they are rendered metallic in the sense that the 3D ESC carried out in these compounds predict that there is no insulating gap. Organic 1D conductors such as TTF-TCNQ are examples of Tomonaga-Luttinger liquids, where the fundamental excitations are not electron-like quasiparticles but gapless and fractionalized excitations of spin and charge. ARPES measurements carried out on TTF-TCNQ accordingly find that the lowest energy excitations are spinons and holons~\cite{claessen2002} and not the electronic quasiparticles of the Fermi liquid state expected in 3D metals. Multiple bands may be important in \TMB\ and this raises the possibility that there may be multiple Fermi surfaces for the spinons and holons, and as well quasiparticles, depending on the strength of interband coupling.

The role of interactions in 1D systems remains a matter of debate. Residual interactions in the Tomonaga-Luttinger state of the 1D organic conductors are sufficient to drive spin-density wave order, Peierls instabilities and even unconventional superconductivity~\cite{ardavan2012,jerome2012}. Long-ranged order necessarily results from interchain coupling, but we are interested mainly in the possibility that these interactions could be sufficiently strong and sufficiently short ranged that Mott-like physics can be realized in 1D systems, where the application of non-thermal variables like pressure, doping, or magnetic field can drive zero temperature electronic instabilities that are analogous to the Mott-Hubbard transition found in 2D~\cite{kanoda2011} and 3D~\cite{kotliar2004} systems. In 3D systems, the stabilization of moments in metals is described within two different paradigms and we ask whether aspects of these scenarios are appropriate for understanding interacting 1D systems in general, and \TMB\ in particular. In the delocalized case, magnetic moments are induced by interactions in a band of delocalized states, and the moment bearing electrons continue to be included in the Fermi surface. The results reported here suggest that the magnetic character of \TMB\ can be understood in this way, although we concede that accurately reproducing local interactions is notoriously difficult in ESC. If this conclusion is correct, then \TMB\ is a more strongly interacting analog of TTF-TCNQ and its 1D relatives. In the localized case, the Coulomb interactions are strong enough to localize moment-bearing electrons, which are not contained in the Fermi surface. The coexistence of a 1D chain of moments with spinon and holon excitations, coexisting with a separate Fermi surface associated with the conduction electrons would be a more exotic outcome for \TMB, likely to already have been realized in the metallic spin chain compound Yb$_{2}$Pt$_{2}$Pb~\cite{Wu-2016}. It is fair to say that it is uncertain at this point whether the Mn $d$-electron states responsible for the magnetism in \TMB\ are localized or delocalized, how many Fermi surfaces are present and of what type (spinon, holon, or quasiparticle). Although the crystal structure of \TMB\ with its Mn-chains is evocative, it is only the observation of fractionalized excitations from ARPES, Resonant Inelastic X-Ray Scattering or inelastic neutron scattering measurements that will determine whether \TMB\ is truly a 1D system. In any case, it is certain that \TMB\ is a unique system that demands further exploration, possibly as a bridge between localized moment spin chains and one-dimensional metals with correlations.\\

\section{Conclusion}
We have investigated the structural, magnetic, electrical transport and thermal properties of a previously unexplored compound \TMB\ that contains 1D chains of Mn-spins running along the crystallographic $c$-axis. Our experimental results, along with the results of electronic structure calculations, establish that this material has a  ground state with moderately strong electronic correlations, leading to Fermi liquid behavior at low temperatures. AFM order with small staggered moments is observed at 1.9 K. Our results suggest that \TMB\ is naturally tuned to the vicinity of a QCP, which may correspond to the emergence of the Mn moment from a correlated band from the onset of AFM order that is driven by interactions among these Mn chains, or from both types of instability. Proximity to these QCPs may explain why the underlying magnetic state in \TMB\ is fragile, with spatial and temporal fluctuations nearly overwhelming the ordering of the Mn moments. Future investigations of these putative QCPs for \TMB\ using pressure or doping will be valuable for establishing the appropriateness of this scenario, and indeed in testing whether the 1D character of \TMB\ is robust, leading to fractionalized excitations such as spinons and holons.\\

{\setlength{\parindent}{1in}{\bf Acknowledgments}}\\

We thank O. Prakash, W. J. Gannon, G. A. Sawatzky, I. Affleck and D. C. Johnston for helpful discussions.  The work at Texas A$\&$M University (AP, PM, MK, HH and MA) was supported by National Science Foundation through the grant NSF-DMR-1807451. HW and XQ acknowledge support from NSF under award DMR-1753054. Portions of this research were conducted with the advanced computing resources provided by Texas A $\&$ M High Performance Research Computing. The identification of any commercial product or trade name does not imply endorsement or recommendation by the National Institute of Standards and Technology.


\begin{thebibliography}{99}

\bibitem{Giamarchi-2004} T. Giamarchi, Quantum physics in one dimensions, (Oxford University Press Inc., New York 2004).

\bibitem{lake2009} B. Lake, A. M. Tsvelik, S. Notbohm, D. A. Tennant, T. G. Perring, M. Reehuis, C. Sekar, G. Krabbes,    and B. Büchner, Confinement of fractional quantum number particles in a condensed-matter system. Nat. Phys. $\bf{6}$, 50 (2009).

\bibitem{mourigal2013} M. Mourigal, M. Enderle, A. Klopperpieper, J. S. Caux, A. Stunault, and H. M. Ronnow, Fractional spinon excitation sin the quantum Heisenberg antiferromagnetic chain, Nat. Phys. $\bf{9}$, 435 (2013).

\bibitem{schlappa2018} J. Schlappa, U. Kumar, K. J. Zhou, S. Singh, M. Mourigal, V. N. Strocov, A. Revcolevschi, L. Patthey, H. M. Ronnow, S. Johnston, and T. Schmitt, Probing multi-spin excitations outside fot he two-spinon continuum in the antiferromagnetic spin chain cuprate Sr$_{2}$CuO$_{3}$, Nat. Commun. $\bf{9}$, 5394 (2018).

\bibitem{Hirakawa-1967} S. Kadota, I. Yamada, S. Yoneyama, and K. Hirakawa, Formation of one-dimensional antiferromagnet in KCuF$_3$ with the perovskite structure, J. Phys. Soc. Jpn. $\bf{23}$, 751 (1967).

\bibitem{Lake-2005} B. Lake, D. A. Tennant, C. D. Frost, and S. E. Nagler, Quantum criticality and universal scaling of a quantum antiferromagnet, Nat. Mat. $\bf{4}$, 329 (2005).

\bibitem{Lee-2012} J. C. T. Lee, S. Yuan, S. Lal, Y. I. Joe, Y. Gan, S. Smadici, K. Finkelstein, Y. Feng, A. Rusydi, P. M. Goldbart, S. L. Cooper, and P. Abbamonte, Two-stage orbital order and dynamical spin frustration in KCuF$_3$, Nat. Phys. $\bf{8}$, 63 (2012).

\bibitem{Pavarini-2008} E. Pavarini, E. Koch, and A. I. Lichtenstein, Mechanism for orbital ordering in KCuF$_3$, Phys. Rev. Lett. $\bf{101}$, 266405 (2008).

\bibitem{Kim-1996} C. Kim, A. Y. Matsuura, Z.-X. Shen, N. Motoyama, H. Eisaki, S. Uchida, T. Tohyama, and S. Maekawa, Observation of spin-charge separation in one-dimensional SrCuO$_2$, Phys. Rev. Lett. $\bf{77}$, 4054 (1996).

\bibitem{Schlappa-2012} J. Schlappa, K. Wohlfeld, K. J. Zhou,	M. Mourigal, M. W. Haverkort,	V. N. Strocov, L. Hozoi, C. Monney, S. Nishimoto,	S. Singh,	A. Revcolevschi, J.-S. Caux, L. Patthey, H. M. R{\o}nnow, J. van den Brink, and T. Schmitt,
Spin–orbital separation in the quasi-one-dimensional Mott insulator Sr$_2$CuO$_3$, Nature $\bf{485}$, 82 (2012).
		
\bibitem{Wu-2016} L. S. Wu, W. J. Gannon, I. A. Zaliznyak, A. M. Tsvelik, M. Brockmann, J.-S. Caux, M. S. Kim, Y. Qiu, J. R. D. Copley, G. Ehlers, A. Podlesnyak, and M. C. Aronson, Orbital-exchange and fractional quantum number excitations in an f-electron metal, Yb$_2$Pt$_2$Pb, Science $\bf{352}$, 1206 (2016).

\bibitem{Jompol-2009} Y. Jompol, C. J. B. Ford, J. P. Griffiths, I. Farrer, G. A. C. Jones, D. Anderson, D. A. Ritchie, T. W. Silk, and A. J. Schofield, Probing spin-charge separation in a Tomonaga-Luttinger liquid, Science $\bf{325}$, 597 (2009).

\bibitem{kanoda2011} K. Kanoda and  R. Kato, Mott physics in organic conductors with triangular lattices, Annu. Rev. Condens. Matter Phys. $\bf{2}$, 167-188(2011).

\bibitem{jerome2012} D. Jerome, Organic superconductors: when correlations and magnetism walk in, J. Supercond. Nov. Mat. $\bf{25}$, 633-655 (2012).

\bibitem{ardavan2012} A. Ardavan, S. Brown, A. Kagoshima, K. Kanoda, K. Kuroki, H. Mori, M. Ogata, S. Uji, and J. Wosnitza, Recent topics of organic superconductors, J. Phys. Soc. Jpn. $\bf{81}$, 011004 (2012).

\bibitem{claessen2002} R. Claessen, M. Sing, U. Schwingenschlogl, P. Blaha, M. Dressel, and C. S. Jacobsen, Spectroscopic signatures of spin-charge separation in the quasi-one-dimensional organic conductor TTF-TCNQ, Phys. Rev. Lett. $\bf{88}$, 096402 (2002).

\bibitem{sikkema1997} A. E. Sikkema, I. Affleck, and S. R. White, Spin gap in a doped chain, Phys. Rev. Lett. $\bf{79}$, 929 (1997).

\bibitem{tsunetsugu1997} H. Tsunetsugu and M. Sigrist, The ground-state phase diagram of the one-dimensional Kondo lattice model, Rev. Mod. Phys. $\bf{69}$, 809-863 (1997).

\bibitem{tsvelik2015} A. M. Tsvelik and O. M. Yevtushenko, Quantum Phase Transition and Protected Ideal Transport in a Kondo Chain, Phys. Rev. Lett. $\bf{115}$, 216402 (2015).

\bibitem{schimmel2016} D. J. Schimmel, A. M. Tsvelik, and O. M. Yevtushenko, Low energy properties of the Kondo chain in the RKKY regime, New. J. Phys. $\bf{18}$, 053004 (2016).

\bibitem{khait2018} I. Khait, P. Azaria, C. Hubig, U. Schollwock, and A. Auerbach, Doped Kondo chain, a heavy Luttinger liquid, Proc. Nat. Acad. Sci. $\bf{115}$, 5140 (2018).

\bibitem{Richter-1997} C. G. Richter, W. Jeitschko, B. K\"unnen, and M. H. Gerdes, The Ternary Titanium Transition Metal Bismuthides Ti$_{4}T$Bi$_2$ with $T =$ Cr, Mn, Fe, Co, and Ni, J. Solid State Chem., $\bf{133}$, 400 (1997).

\bibitem{Rytz-1999} R. Rytz and R. Hoffmann, Chemical bonding in the ternary transition metal bismuthides Ti$_{4}T$Bi$_2$ with $T =$ Cr, Mn, Fe, Co, and Ni, Inorg. Chem. $\bf{38}$, 1609 (1999).

\bibitem{Carvajal-1993} J. Rodr\'iguez-Carvajal, Recent advances in magnetic structure determination by neutron powder diffraction, Physica B $\bf{192}$, 55 (1993); see also www.ill.eu/sites/fullprof/.

\bibitem{lynn2012} J. W. Lynn, Y. Chen, S. Chang, Y. Zhao, S. Chi, W. Ratcliff, B. G. Ueland, and R. W. Erwin, Double focusing thermal triple axis spectrometer at the NCNR, J. Research NIST $\bf{117}$, 61 (2012).

\bibitem{Oishi-2009} R. Oishi, M. Yonemura, Y. Nishimaki, S. Torii, a. Hoshikawa, T. Ishigaki, T. Morishima, K. Mori, T. Kamiyama, Rietveld analysis software for J-PARC, Nucl. Instr. Meth. Phys. Res. A $\bf{600}$, 94 (2009).

\bibitem{Oishi-2012} R. Oishi-Tomiyasu, M. Yonemura, T. Morishima, a. Hoshikawa, S. Torii, T. Ishigaki, T. Kamiyama, Application of matrix decomposition algorithms for singular matrices to the Pawley method in Z-Rietveld, J. Appl. Cryst. $\bf{45}$, 299 (2012).

\bibitem{hohenberg1964} P. Hohenberg and W. Kohn, Inhomogeneous electron gas, Phys. Rev. B $\bf{136}$, 864 (1964).

\bibitem{kohn1965} W. Kohn and L. J. Sham, Self-consistent equations including exchange and correlation effects, Phys. Rev. $\bf{140}$, A1133  (1965).

\bibitem{kresse1996} G. Kresse and J. Furthmüller, Efficient iterative schemes for ab initio total-energy calculations using a plane-wave basis set, Phys. Rev. B $\bf{54}$, 11169 (1996).

\bibitem{blochl1994} P. E. Blochl, Projector augmented-wave method, Phys. Rev. B $\bf{50}$, 17953 (1994).

\bibitem{perdew1996} J. P. Perdew, K. Burke, and M. Ernzerhof, Generalized gradient approximation made simple, Phys. Rev. Lett. $\bf{77}$, 3865 (1996).

\bibitem{klimes2010} J. Klime$\breve{s}$ J. Klime\v{s}, D. R. Bowler, and A. Michaelides, Chemical accuracy for the van der Waals density functional, J. Phys. Condens. Matter $\bf{22}$, 022201 (2010).

\bibitem{dudarev1998} S. L. Dudarev, G. A. Botton, S. Y. Savrasov, C. J. Humphreys, and A. P. Sutton, Electron-energy-loss spectra and the structural stability of nickel oxide: An LSDA+U study, Phys. Rev. B $\bf{57}$, 1505 (1998).

\bibitem{jacko2009} A. C. Jacko, J. O. Fj{\ae}restad, and B. J. Powell, A unified explanation of the Kadowaki–Woods ratio in
strongly correlated metals, Nat. Phys. $\bf{5}$, 422 (2009).
		
\bibitem{marzari2012} N. Marzari, A. A. Mostofi, J. R. Yates, I. Souza, and D. Vanderbilt, Maximally localized Wannier functions: Theory and applications, Rev. Mod. Phys. $\bf{84}$, 1419  (2012).

\bibitem{qian2008} X. Qian, J. Li, L. Qi, C. Z. Wang, T. L. Chan, Y. X. Yao, K. M. Ho, and S. Yip, Quasiatomic orbitals for ab initio tight-binding analysis, Phys. Rev. B $\bf{78}$, 245112  (2008).

\bibitem{yin2011} Y. P. Yin, K. Haule, and G. Kotliar, Kinetic frustration and the nature of the magnetic and paramagnetic states in iron pnictides and iron chalcogenides, Nat. Mater. $\bf{10}$, 932 (2011).

\bibitem{mcnally2015} D. E. McNally, S. Zellman, Z. P. Yin, K. W. Post, Hua He, K. Hao, G. Kotliar, D. Basov, C. C. Homes, and M. C. Aronson, From Hunds insulator to Fermi liquid: Optical spectroscopy study of K doping in BaMn$_2$As$_2$, Phys. Rev. B $\bf{92}$, 115142 (2015).

\bibitem{Sangeetha-2016} N. S. Sangeetha, A. Pandey, Z. A. Benson, and D. C. Johnston, Antiferromagnetism in trigonal SrMn$_{2}$AS$_{2}$ and CaMn$_{2}$As$_{2}$ single crystals. Phys. Rev. B $\bf{94}$, 094417 (2016).

\bibitem{mcnally2014} D. E. McNally, J. W. Simonson, K. W. Post, Z. P. Yin, M. Pezzoli, G. J. Smith, V. Leyva, C. Marques, L. DeBeer-Schmitt, A. I. Kolesnikov, Y. Zhao, J. W. Lynn, D. N. Basov, G. Kotliar, and M. C. Aronson, Origin of the charge gap in LaMnPO, Phys. Rev. B $\bf{90}$, 180403(R)(2014).

\bibitem{WeissTemp-BF} A Wiess temperature $\theta_{\rm c}$ was used as a fit parameter in the fitting of $M(H)$ data by the BF. However, since the best fit results in $\theta_{\rm c} = 0$ within the error bars, the $\theta_{\rm c}$ was removed from the fit expression during the final fitting. 

\bibitem{Johnston-2012} D. C. Johnston, Magnetic Susceptibility of Collinear and Noncollinear Heisenberg Antiferromagnets, Phys. Rev. Lett. {\bf 109}, 077201 (2012).

\bibitem{Goetsch-2012} R. J. Goetsch, V. K. Anand, A. Pandey, and D. C. Johnston, Structural, thermal, magnetic, and electronic transport properties of the LaNi$_2$(Ge$_{1−x}$P$_x$)2 system, Phys. Rev. B {\bf 85}, 054517 (2012).

\bibitem{wills2000} A. S. Wills, A new protocol for the determinations of magnetic structures using simulated annealing and
 representational analysis (SARAh), Physica B $\bf{287}$, 680 (2000).

\bibitem{simonson2012} J. W. Simonson, G. J. Smith, K. Post, M. Pezzoli, J. J. Kistner-Morris, D. E. McNally, J. E. Hassinger, C. S. Nelson, G. Kotliar, D. N. Basov, and M. C. Aronson, Magnetic and structural phase diagram of CaMn$_{2}$Sb$_{2}$, Phys. Rev. B $\bf{86}$, 184430 (2012).

\bibitem{Ann-2009} J. An, A. S. Sefat, D. J. Singh, and M.-H. Du, Electronic structure and magnetism in BaMn$_{2}$As$_{2}$ and BaMn$_{2}$Sb$_{2}$, Phys. Rev. B $\bf{79}$, 075120 (2009).

\bibitem{Singh-2009} Y. Singh and D. C. Johnston, Magnetic, transport, and thermal properties of single crystals of the layered arsenide 
BaMn$_{2}$As$_{2}$, Phys. Rev. B $\bf{79}$, 094519 (2009).

\bibitem{okuda1999} H. Okuda, S. Senba, H. Sato, K. Shimada, H. Namatame, and M. Taniguchi, Electronic structure of MnSb and MnP, J. Electron Spectrosc.  $\bf{101-103}$, 657 (1999).

\bibitem{shanavas2014} K. V. Shanavas, D. Parker, and D. J. Singh, Theoretical study on the role of dynamics on the unusual magnetic properties in MnBi, Sci. Rep. $\bf{4}$, 7222 (2014).

\bibitem{okabayashi2004} J. Okabayashi, M. Mizuguchi, K. Ono, M. Oshima, A. Fujimori, H. Kuramochi, and H. Akinaga, Density-dependent electronic structure of zinc-blende-type MnAs dots on GaAs(001) studied by in situ photoemission spectroscopy, Phys. Rev. B $\bf{70}$, 233305 (2004).

\bibitem{Pandey-2015}A. Pandey and D. C. Johnston, Ba$_{0.4}$Rb$_{0.6}$Mn$_{2}$As$_{2}$: a prototype half-metallic ferromagnet. Phys. Rev. B $\bf{92}$, 174401 (2015).

\bibitem{carbone2006} F. Carbone, M. Zangrando, A. Brinkman, A. Nicolaou, F. Bondino, E. Magnano, A. A. Nugroho, F. Parmigiani, Th. Jarlborg, and D. van der Marel, Electronic structure of MnSi: the role of electron-electron interactions, Phys. Rev. B $\bf{73}$, 085114 (2006).

\bibitem{yamaoka2009}H. Yamaoka, N. Tsujii, I. Jarrige, Y. Takahashi, J. Chaboy, H. Oohashi, K. Handa, J. Ide, T. Tochio, Y. Ito, T. Uruga, and H. Yoshikawa, Electronic structure of YMn$_{2}$ and Y$_{0.96}$Lu$_{0.04}$Mn$_{2}$ studied by x-ray emission spectroscopy, Phys. Rev. B $\bf{80}$, 115110 (2009).

\bibitem{marques2011}C. Marques, Y. Janssen, M. S. Kim, L. Wu, S. X. Chi, J. W. Lynn, and M. C. Aronson, HfFeGa$_{2}$ and
HfMnGa$_{2}$: Transition-metal-based itinerant ferromagnets with low Curie temperatures, Phys. Rev. B $\bf{83}$, 184435 (2011).

\bibitem{coey2010} J. M. D. Coey, \it{Magnetism and Magnetic Materials}\rm, Cambridge University Press (2010).

\bibitem{kotliar2004} G. Kotliar and D. Vollhardt. Strongly correlated materials: insights from dynamical mean field theory, Phys. Today $\bf{57(3)}$, 53 (2004).

\end{thebibliography}
\end{document}